\documentclass[preprint,12pt]{elsarticle}

\usepackage{amssymb}

\usepackage[dvipsnames]{xcolor}
\usepackage{subcaption}

\usepackage{multirow}
\usepackage{longtable}

\journal{Current Opinion in Structural Biology}

\begin{document}

\begin{frontmatter}



\title{Deep learning for reconstructing protein structures from cryo-EM density maps: recent advances and future directions}


\affiliation[label1]{organization={Department of Electrical Engineering and Computer Science}, addressline={University of Missouri}, city={Columbia}, postcode={65211}, state={Missouri}, country={USA}}

\author{Nabin Giri\fnref{label1}}
\ead{ngzvh@missouri.edu}
\author{Raj S. Roy\fnref{label1}}
\ead{rsr3gt@missouri.edu}
\author{Jianlin Cheng\corref{cor1}\fnref{label1}}
\ead{chengji@missouri.edu}
\ead[url]{http://calla.rnet.missouri.edu/cheng/}
\cortext[cor1]{Corresponding author}

\begin{abstract}
Cryo-Electron Microscopy (cryo-EM) has emerged as a key technology to determine the structure of proteins, particularly large protein complexes and assemblies in recent years. A key challenge in cryo-EM data analysis is to automatically reconstruct accurate protein structures from cryo-EM density maps. 
In this review, we briefly overview various deep learning methods for building protein structures from cryo-EM density maps, analyze their impact,
and discuss the challenges of preparing high-quality data sets for training deep learning models. Looking into the future, more advanced deep learning models of effectively integrating cryo-EM data with other sources of complementary data such as protein sequences and AlphaFold-predicted structures need to be developed to further advance the field.  

\end{abstract}

\begin{graphicalabstract}
\begin{figure}
    \centering
    \includegraphics[width=\textwidth]{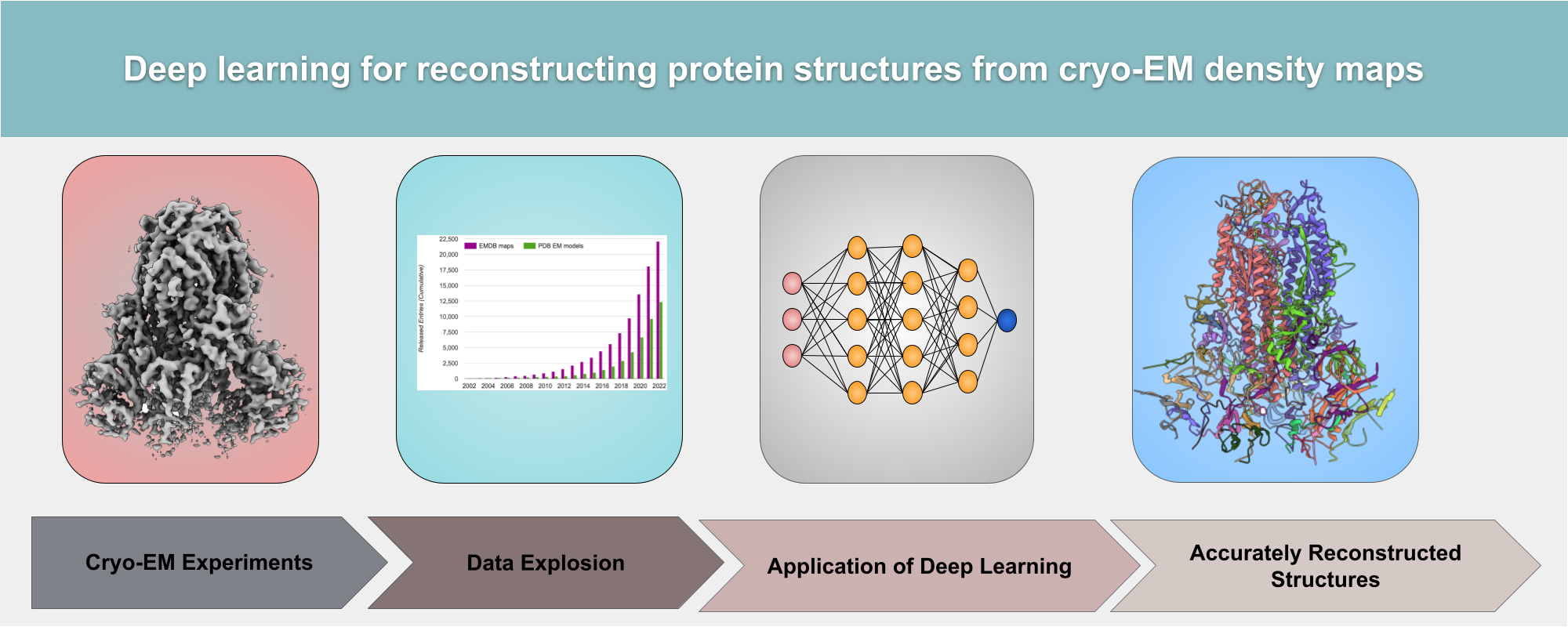}
    \label{fig:em_growth}
\end{figure}
\end{graphicalabstract}

\begin{highlights}
\item Deep learning is a promising technique for efficient, automatic, and accurate reconstruction of protein structures from cryo-EM density maps
\item Advanced convolutional neural networks and U-Nets have been successfully applied to reconstruct protein structures from high-resolution cryo-EM density maps 
\item Creating high-quality cryo-EM data sets for training and testing deep learning methods is important and there is a significant need of curating such data sets to facilitate the development of deep learning methods
\item Better structure reconstruction can be obtained by combining AlphaFold predicted structure models and cryo-EM data and by integrating cryo-EM based structure determination techniques and protein structure prediction techniques. 
\item More advanced deep learning architectures and better integration of multiple sources of complementary data are needed to advance the field
\end{highlights}

\begin{keyword}
cryo-electron microscopy (cryo-EM) \sep protein structure  \sep machine learning \sep deep learning


\end{keyword}

\end{frontmatter}


\section{Introduction}
\label{sec:intro}
\begin{figure}
     \centering
     \begin{subfigure}[b]{0.47\textwidth}
         \centering
         \includegraphics[width=\textwidth]{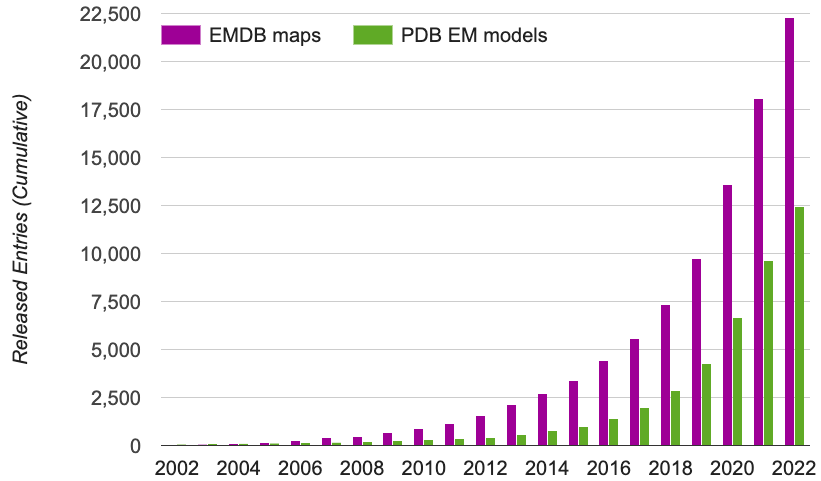}
         \caption{Growth of cryo-EM density maps and corresponding protein structures}
         \label{fig:em_growth}
     \end{subfigure}
     \hfill
     \begin{subfigure}[b]{0.43\textwidth}
         \centering
         \includegraphics[width=\textwidth]{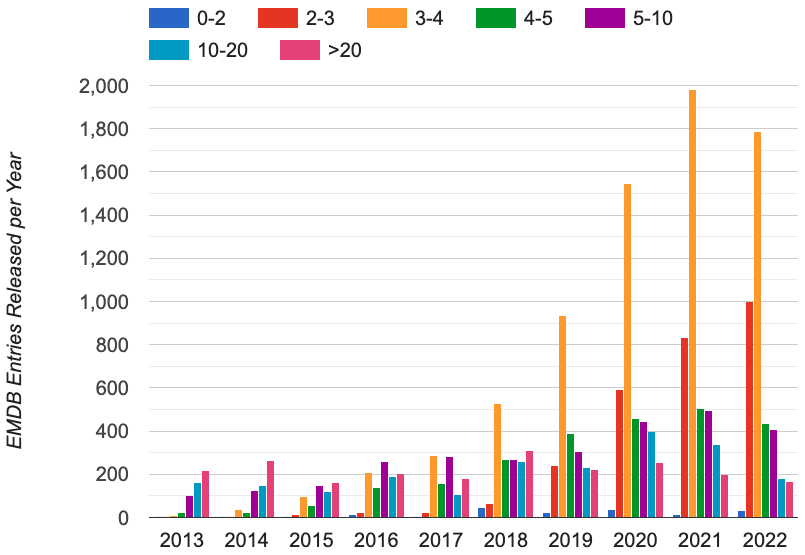}
         \caption{The distribution of the resolution of the cryo-EM density maps ($\AA$)}
         \label{fig:em_reso}
     \end{subfigure}
        \caption{The growth of cryo-EM density maps and cryo-EM-derived protein structures and the distribution of the resolution of the density maps. The statistics was obtained from EMDataResource \cite{EMDB}, an unified data resource for 3-Dimension electron microscopy (3DEM) on 2022-09-14.}
        \label{fig:emdr_fig}
\end{figure}

Cryo-EM is revolutionizing structural biology due to its unique capability of determining the structures of large protein complexes and assemblies. 
The atomic-resolution structure determination for proteins enabled by cryogenic electron microscopy (cryo-EM) \cite{resolution_revolution}, allows us to understand the complex biological processes carried out by proteins as well as to identify potential therapeutic protein targets for drug discovery. However, reconstructing $\textit{de novo}$ protein structures from high-resolution ($\sim$ 3 - 4 $\AA$) cryo-EM density maps, which accounts for a large portion of cryo-EM density maps deposited currently in the EMDB \cite{EMDB}, is time-consuming and challenging when homologous template structures for target proteins are not available. For instance, as shown in \textbf{Figure \ref{fig:emdr_fig}}, in the current year 2022, only about 12,500 out of 22,300 density maps of high-resolutions deposited to EMDB have a complete atomic structure available in Protein Data Bank (PDB) \cite{PDB}.

Accurately reconstructing protein structures from cryo-EM maps is a challenging process because the data is often noisy and incomplete and target protein structures can be large and complex. Traditional methods based on energy optimization such as EM-Fold \cite{em_fold}, Gorgon \cite{gorgon}, Rosetta \cite{rosetta}, Pathwalking \cite{pathwalking}, MAINMAST \cite{mainmastseg, mainmast}, VESPER \cite{vesper}, and Phenix \cite{phenix} have made valuable progress in reconstructing protein structures from cryo-EM density maps. These methods rely on extensive physics-based or statistical potential-based optimization algorithms that require high computational resources. These methods often need manual intervention and trials to extract features from the cryo-EM density maps to obtain accurate reconstruction of protein structure. 

A different strategy to automatically determine protein structures from cryo-EM density maps is to use the data-driven machine learning approach \cite{machine_learning}, a kind of artificial intelligence (AI) technology,  to directly learn a mapping from cryo-EM density maps to protein structures from the large amount of known cryo-EM data and their corresponding protein structures (i.e., labels). Early AI methods in the field are based on shallow machine learning techniques such as k-nearest neighbor, support-vector machines, or k-means clustering techniques. These methods such as RENNSH \cite{rennsh}, SSELearner \cite{sselearner}, and Pathwalking \cite{pathwalking} are able to identify only secondary structures or simplified backbone structures and often are unable to achieve the optimal solution.

\begin{figure}
    \centering
    \includegraphics[width=\textwidth]{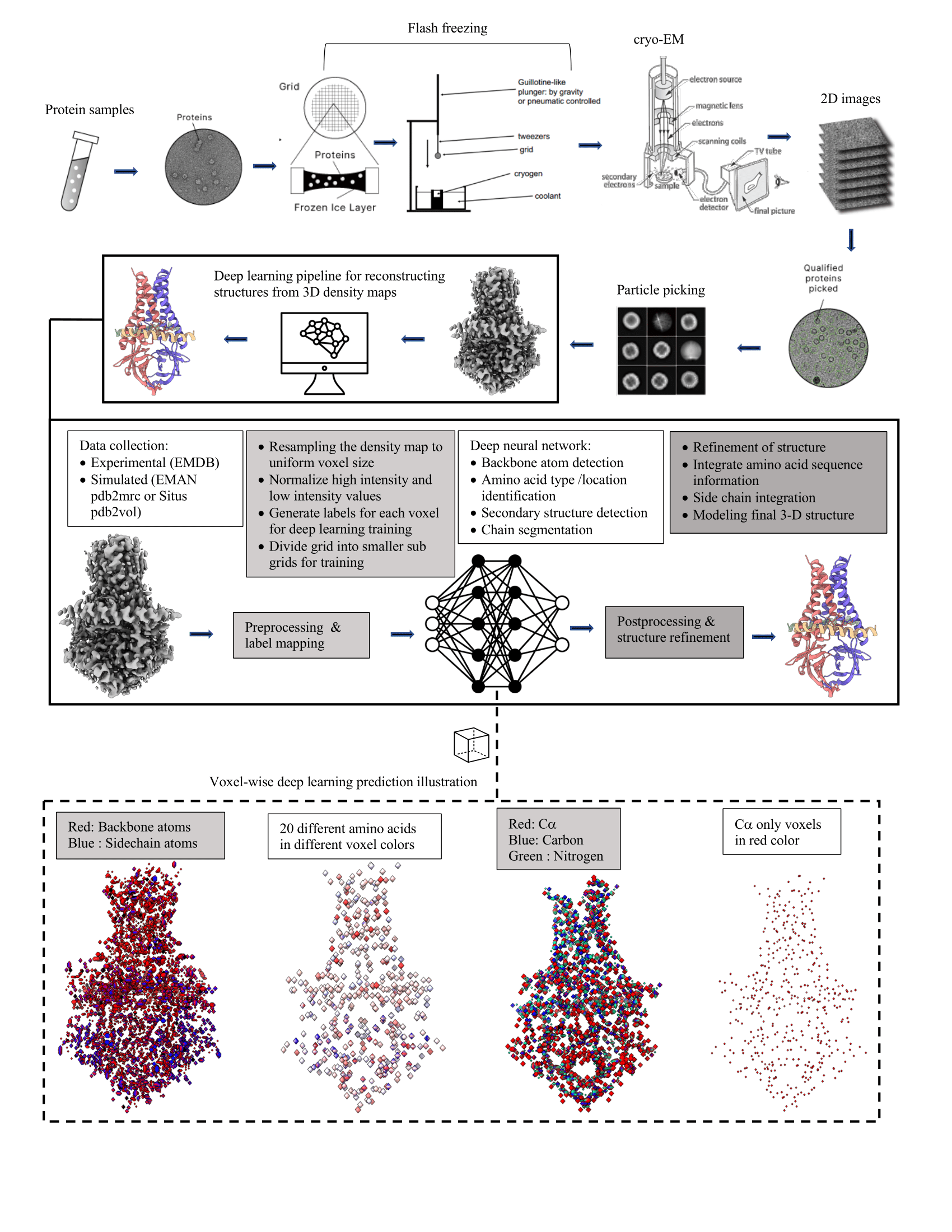}
    \caption{A summary of a cryo-EM density map generation and protein structure reconstruction pipeline powered by deep learning. The density map (EMD-22898) illustrated in the figure is for SARS-CoV-2 ORF3a \cite{orf3a}. PDB ID: 7KJR.}
    \label{fig:summary_cryo}
\end{figure}

To overcome the challenges of the traditional optimization methods and early machine learning methods, deep learning methods \cite{deep_learning_medical} have been developed to automatically reconstruct three-dimensional (3D) protein structures from cryo-EM density maps with significant success in recent years (see \textbf{Figure \ref{fig:summary_cryo}} for a summary of a general cryo-EM protein structure determination pipeline powered by deep learning). In this article, we review the recent development of deep learning technology in the field, analyze their impacts, investigate the challenging issues in preparing data to train deep learning models, and discuss some new trends to further advance the field.

\section{Deep learning reconstruction of protein structures from cryo-EM density maps}
\label{sec:deep-learning}
Deep learning, also called deep neural network, is currently the most powerful machine learning method of predicting the properties of an object from the input data describing the object. It has achieved great success in many fields including a recent major breakthrough in predicting protein structure from sequence by AlphaFold \cite{alphafold}. 
Compared to other machine learning methods, deep learning has a unique capability of extracting informative features for pattern recognition from raw data automatically, making it suitable for reconstructing protein structures from raw density maps in which only a large amount of numbers rather than informative features are available. 

It is worth noting that deep learning has been applied to almost all the areas of cryo-EM data analysis \cite{cryo_review, cryogan, cryo_drgn, e2gmm, CDAE, map_denoising, deepcryopicker} from sample preparation, particle picking, density map denoising, and to the final step of 3-D structure determination. Due to the space limit, this review is focused on the last step of cryo-EM data analysis - reconstructing protein structures from density maps. The deep learning architectures designed for this task and how to prepare data to train them are discussed in the two subsections below.

\subsection{Deep learning architectures for reconstructing protein structures from cryo-EM density maps}

\begin{figure}
     \centering
     \hfill
     \begin{subfigure}[b]{0.32\textwidth}
         \centering
         \includegraphics[width=\textwidth]{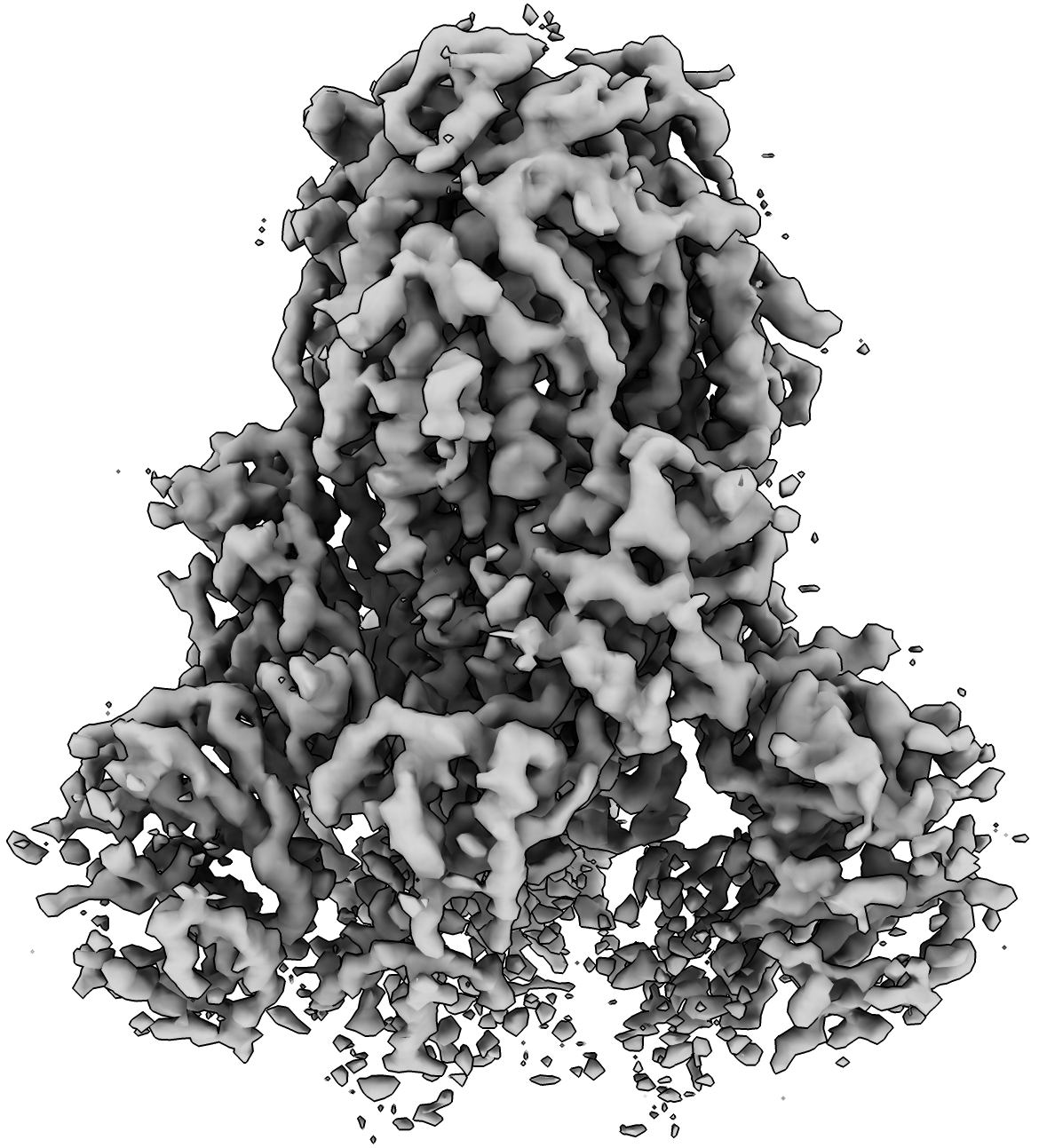}
         \caption{ }
         \label{fig:cryo_6732}
     \end{subfigure}
     \begin{subfigure}[b]{0.32\textwidth}
         \centering
         \includegraphics[width=\textwidth]{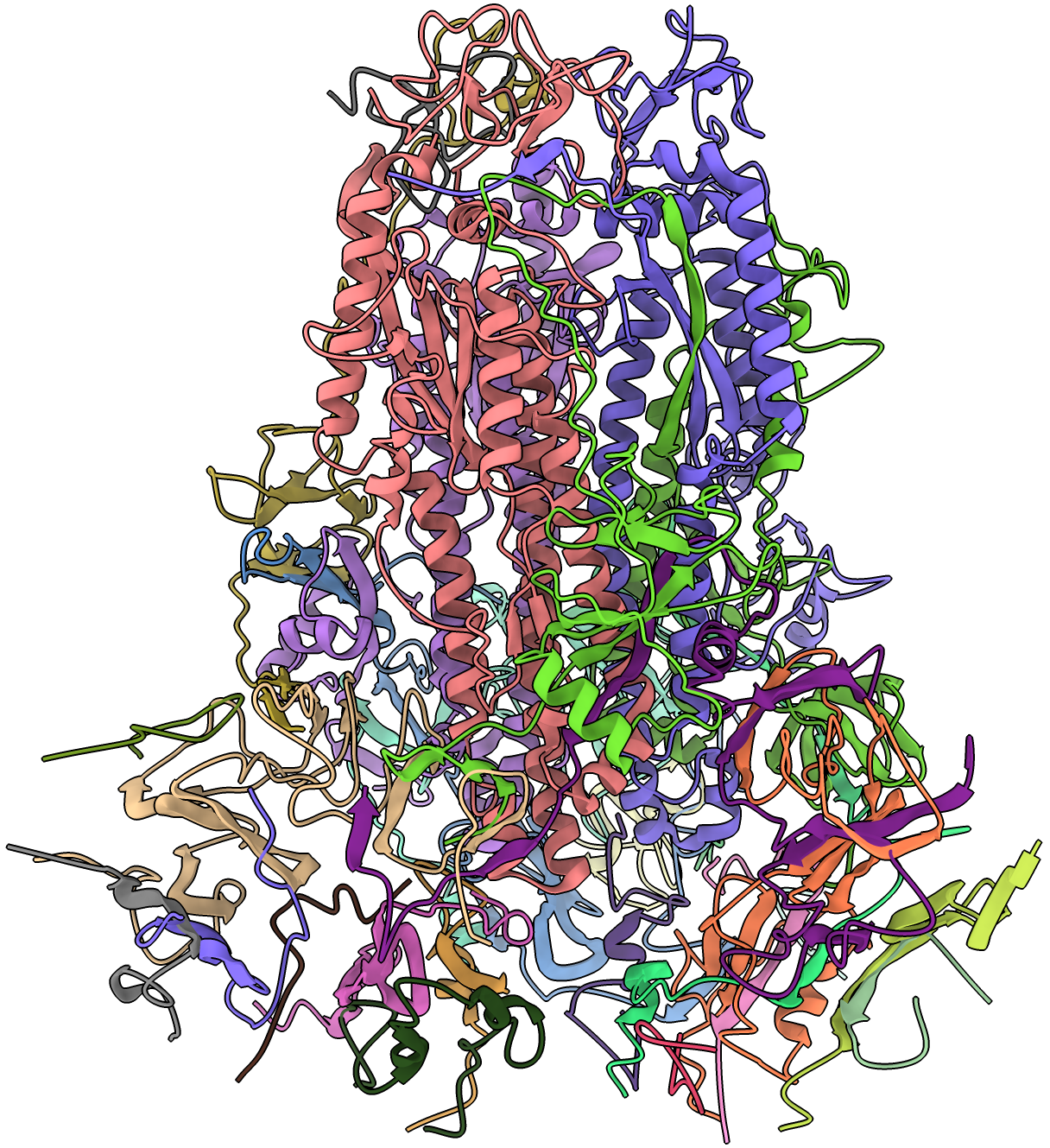}
         \caption{}
         \label{fig:pdb_6732}
     \end{subfigure}
     \begin{subfigure}[b]{0.32\textwidth}
         \centering
         \includegraphics[width=\textwidth]{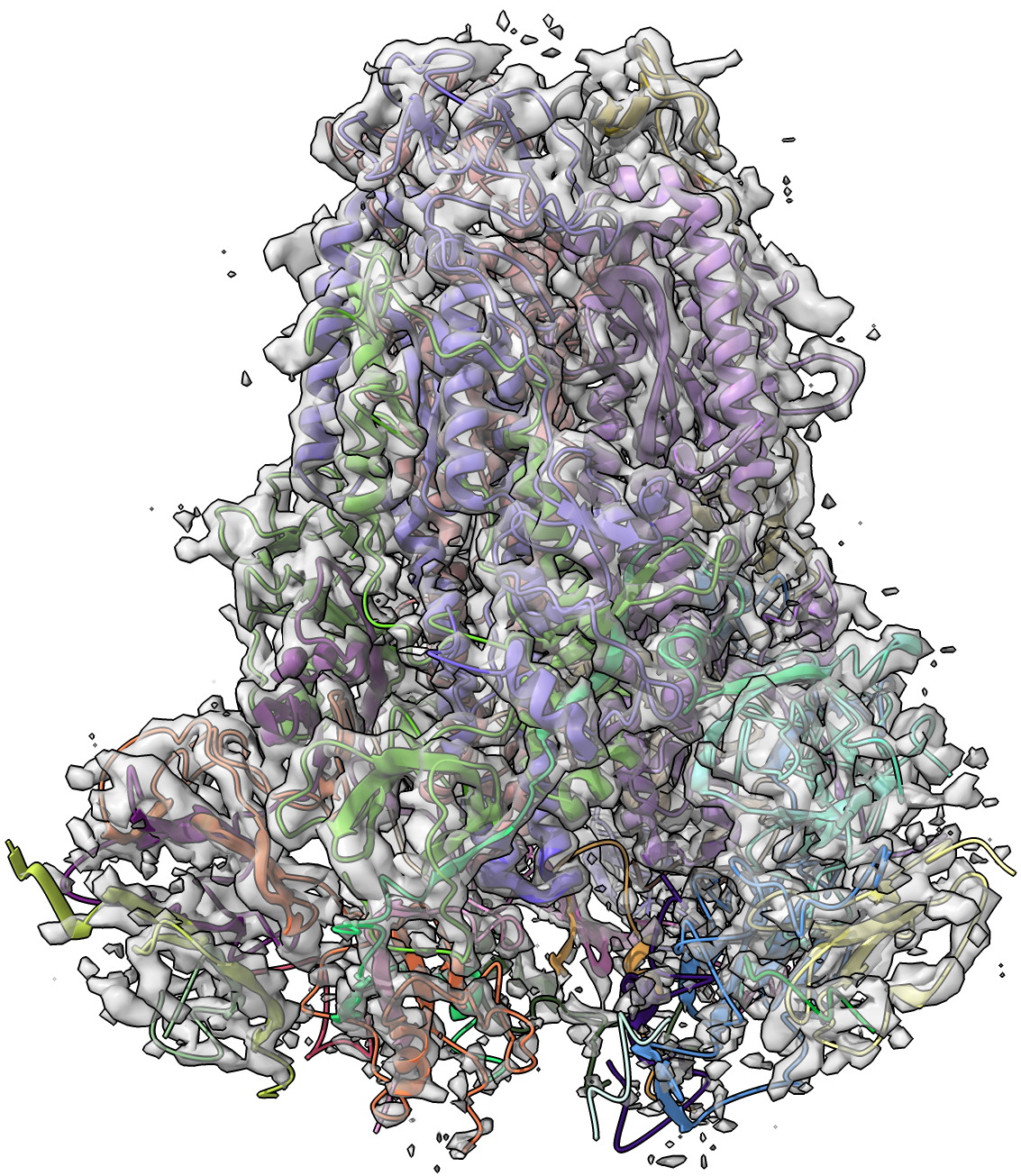}
         \caption{}
         \label{fig:map_model}
     \end{subfigure}
        \caption{An example of reconstructing a structure from the cryo-EM density map of SARS-CoV spike gycoprotein by deep learning. \textbf{(a)} Density map of SARS-CoV spike glycoprotein \cite{glycoprotein} (EMD-6732) in resolution of 3.8 $\AA$ at recommended contour level of 0.06 (11.0 $\sigma$). \textbf{(b)} The structure reconstructed from EMD-6732 by a deep learning method - DeepTracer. The RMSD is 1.023 $\AA$ with respect to the ground truth structure (PDB ID: 5XLR). \textbf{(c)} The overlay of the density map and reconstructed structure at 0.5 transparency level by UCSF ChimeraX \cite{chimerax}. }
        \label{fig:deeptracer_example}
\end{figure}

Deep learning methods for inferring protein structures from cryo-EM density maps can be classified into different categories based on the neural network architectures (e.g., convolutional neural network (CNN)  \cite{cnn}, U-Net \cite{unet, unet++}, graph convolutional network (GCN) \cite{gcn}, and long- and short-term memory network (LSTM) \cite{lstm} they use and the output (e.g., 3D structure and secondary structure) they generate from density map input.  Early deep learning methods aimed to identity secondary structures from low- and medium-resolution density maps \cite{cnn_based}. As more and more high-resolution density maps became available \cite{resolution_revolution}, recent deep learning methods targeted at directly reconstruct 3D backbone structures (i.e., locations of carbon and nitrogen atoms on the protein backbone) and even full-atom 3D structures (i.e., locations of all/most heavy atoms and amino acid identity/type) from density maps \cite{cascaded_cnn, structure_generator, deeptracer_id, cr_i_tasser, embuild}. An example of deep learning reconstruction of protein structure from cryo-EM density map is showed in \textbf{Figure \ref{fig:deeptracer_example}}.

One of the most widely used deep learning architectures of obtaining protein structural information from density maps is convolution neural network (CNN). CNNs use a mathematical operation known as convolution to extract features from spatially organized data such as a 2D-image or 3D density map to predict the properties of the data (e.g., classifying voxels in a density map into amino acid types). Several CNN methods (mostly 3D-CNN architecture) including Generator \cite{structure_generator}, Emap2sec \cite{emap2sec}, AAnchor \cite{aanchor}, CNN Based \cite{cnn_based}, Cascaded-CNN \cite{cascaded_cnn}, and CR-I-TASSER (mostly 3D CNN) \cite{cr_i_tasser} have been developed to determine secondary structures \cite{emap2sec, cnn_based}, backbone-/full-atom 3D structures \cite{cr_i_tasser, structure_generator, aanchor} or both from cryo-EM density maps \cite{cascaded_cnn}. Cascaded-CNN is the first deep learning de novo method of directly reconstructing 3D structures of proteins from cryo-EM density maps, even though it focuses on building backbone structures. CR-I-TASSER combines the 3D-CNN prediction from cryo-EM maps and an advanced protein structure prediction method - I-TASSER \cite{i_tasser} to build full-atom protein structures. 

Another widely used deep learning architecture in the field is U-Net \cite{unet}, originally designed for biomedical image classification and segmentation tasks. U-Net consists of a series of convolution-based down-sampling layers to condense the input images into smaller dimensions and a series of convolution-based up-sampling counterpart layers to reconstruct the data of the same dimension as in the down-sampling process to classify/segment pixels in the input images. Compared to the standard CNN architectures, U-Nets can be more effectively in extracting multi-level abstract representations of the data through the down-sampling and up-sampling processes. The 2D U-Net architecture has been generalized to 3D U-Net architectures in Haruspex \cite{haruspex} and  EMNUSS \cite{emnuss} to detect secondary structures from cryo-EM density maps and DeepTracer \cite{deeptracer}  and EMBuild \cite{embuild} to reconstruct 3D protein structures from cryo-EM density maps. DeepTracer has been successfully applied to reconstruct the structures of some SARS-CoV proteins from cryo-EM density maps (e.g., \textbf{Figure \ref{fig:deeptracer_example}}). 

In addition to CNN and U-Net, other deep learning architectures such as graph convolutional networks (GCN) and long- and short-term memory network (LSTM) have also been used with CNN to reconstruct protein structures from cryo-EM density maps \cite{structure_generator}. A summary of different deep learning-based methods, their function (e.g., input and output) and availability is presented in \textbf{Table 1}.  \\

\begin{longtable}{|p{2.7cm}||p{2.7cm}|p{6.5cm}|p{1cm}|}
\hline
Methods  & Architecture & Function & Open source\\
\hline
Structure Generator\cite{structure_generator} &  3-D CNN, GCN, Bidirectional LSTM & First use 3-D CNN to identify amino acids and their rotameric identities in an EM map and then GCN and LSTM to \textbf{\textit{build protein structures}}  & $\checkmark$\\
\hline
Emap2sec\cite{emap2sec} & 3-D CNN  & Take voxel cubes as input to \textbf{\textit{identify secondary structures of protein}} &$\checkmark$ \\
\hline
AAnchor\cite{aanchor} & 3-D CNN  & Take in voxel cubes to \textbf{\textit{identify amino acid types and locations}} & $\checkmark$\\
\hline
A CNN Based Method\cite{cnn_based} & 3-D CNN & Take in voxel cubes to \textbf{\textit{detect secondary structures of protein from background}}  & $\times$ \\
\hline
CascadedCNN \cite{cascaded_cnn} & Cascaded 3-D CNN & Take in voxel cubes to \textbf{\textit{identify C$\alpha$ atoms of protein backbone and secondary structures to generate 3D protein structures}} & $\checkmark$\\
\hline
Haruspex\cite{haruspex} & 3-D U-Net & Take in voxel cubes to predict the probabilities of 4 different classes; $\alpha$-helix, $\beta$-sheet, nucleotide, or unassigned to \textbf{\textit{assign secondary structures}}   & $\checkmark$\\
\hline
DeepTracer\cite{deeptracer} & 3-D U-Net & Take in voxel cubes to \textbf{\textit{identify the location of backbone atoms, secondary structures and amino acid types simultaneously to build 3D structure}} & $\checkmark$\\
\hline
DeepTracer ID \cite{deeptracer_id} & DeepTracer (3-D U-Net) and pre-calculated AlphaFold2 protein library   & Use DeepTracer to generate an initial 3D protein structure to search AlphaFold2DB to \textbf{\textit{identify similar structural hits for refinement}} & $\checkmark$\\
\hline
CR-I-TASSER \cite{cr_i_tasser} & 3-D CNN, I-TASSER & Predict C$\alpha$ using 3-D CNN for selecting structural templates for I-TASSER to \textbf{\textit{generate 3D protein structure}} & $\checkmark$ \\
\hline
EMBuild \cite{embuild} & 3-D U-Net++, AlphaFold & Integrate AlphaFold structure prediction, FFT-based global fitting, domain-based semi-flexible refinement, and graph-based iterative assembling with main-chain probability maps predicted by U-Net++ to \textbf{\textit{3D build protein structure}} & $\checkmark$ \\
\hline
EMNUSS \cite{emnuss} & 3-D U-Net++ & Take in voxel cubes to \textbf{\textit{identify secondary structures of protein}} & $\checkmark$ \\
\hline
\caption{Summary of deep learning based methods for protein structure reconstruction from cryo-EM density maps.}\\
\end{longtable}
\label{tab:deep_learning}

Inspired by the recent breakthrough in developing deep learning methods of predicting protein structures from sequences such as AlphaFold \cite{alphafold} and RoseTTAFold \cite{rosettafold}, a new trend is to integrate deep learning methods of reconstruct protein structures from cryo-EM density maps with the advanced computational (e.g., deep learning) methods of predicting protein structures from sequences to obtain more accurate structural models. For instance, DeepTracer ID \cite{deeptracer_id} first uses DeepTracer to build an initial structure from cryo-EM density maps and then search the structure against a database of AlphaFold-predicted structures to identify similar structural hits to enhance the reconstructed structure. EMBuild \cite{embuild} combines the structures reconstructed from cryo-EM maps, AlphaFold-predicted structural models and other protein structural refinement methods to construct accurate structures for protein complexes. DeepProLigand \cite{deepproligand} integrates the protein structural models reconstructed from cryo-EM density maps by DeepTracer with the known template structures containing ligands to model protein-ligand interaction, which was ranked first in the ligand prediction in 2021 EMDataResource Ligand Model Challenge.

\section{Data preparation for training deep learning methods to reconstructing protein structures from cryo-EM density maps}
\label{sec:dataset-preparation}

\subsection{Cryo-EM density map data collection}
Collecting a sufficient amount of high-quality data to train and test deep learning models is critical for any deep learning task. The common way to acquire the \textit{experimental cryo-EM density maps} is through the EMDB \cite{EMDB}. An alternative approach employed by some methods such as Cascaded-CNN \cite{cascaded_cnn} and SSELearner \cite{sselearner} is to simulate the density map from the PDB protein structure. Cascaded-CNN applies \textit{pdb2mrc} from EMAN2 package \cite{eman2}, and VESPER uses \textit{pdb2vol} from Situs package \cite{situs} to generate the \textit{simulated maps}.  However, simulated maps lack complex noise, missing density values, and experimental artifacts which can arise from particle alignment errors, interaction of electron beam with the atoms, or movement of atoms during image capture. Therefore, the deep learning models trained on simulated maps may not work as expected on very noisy experimental data. To address the problem, CR-I-TASSER, EMNUSS and Emap2sec employs a hybrid training approach that uses both simulated maps and experimental maps in the training and validation process.

\subsection{Training data preprocessing}
Prior to using the cryo-EM density map to train deep learning models, it is generally necessary to normalize and standardize the data to make them suitable for machine learning as shown by Cascaded-CNN and DeepTracer, which perform data grid resampling, density value normalization, and grid division. These preprocessing steps ensure the uniformity among density maps and help deep learning models to extract features and recognize patterns more easily. During the grid division, the 3D cryo-EM is splitted into the cubes of a specific size (e.g., 64 $\times$ 64 $\times$ 64 $\AA^{3}$ by Cascaded-CNN and DeepTracer, 50 $\times$ 50 $\times$ 50 $\AA^{3}$ by CR-I-TASSER, 40 $\times$ 40 $\times$ 40 $\AA^{3}$ by Haruspex, and 11 $\times$ 11 $\times$ 11 $\AA^{3}$ by Emap2sec and AAnchor). Each of these cubes is then processed by the deep learning method to classify the voxels into the targeted classes such as amino acid types (identities) and secondary structures.


\section{Future directions}
\label{sec:future}

Deep learning has made a significant impact on protein structure reconstruction from cryo-EM density maps. However, the field is still in the early stage of development. The latest deep learning technology such as graph neural networks \cite{geometric_deeplearning} and attention mechanisms \cite{attention_survey} have not been used in the field.  While CNNs and U-Nets based on convolution are currently the most used methods for structure reconstruction, they have some shortcoming for 3D structural modeling. CNNs are translation-equivariant, but not fully rotation invariant that is desirable for 3D structure analysis. Moreover, the convolution mechanism propagates message in the constrained local receptive field, which is not as effective as the attention mechanism \cite{attention_survey} that can leverage all the input information by automatically weighting the input features according to their relevance as demonstrated by the remarkable success of AlphaFold2 in protein structure prediction. 
More sophisticated deep learning models like attention-based Transformer models \cite{transformer}, 3D-equivariant graph neural networks \cite{se3}, and AlphaFold2-like deep learning models need to be developed to better use cryo-EM data to improve reconstruction accuracy.  

Another important direction is to use deep learning to integrate cryo-EM data with multiple other sources of complementary data such as protein structural models predicted from sequences, structural templates in the Protein Data Bank, and protein sequences to more accurately reconstruct protein structures from noisy density maps that often miss the density values of some atoms.  The current integration process is limited to shallow data combination. For instance, DeepTracer ID uses AlphaFold models to refine the structural models predicted from structural models reconstructed from deep learning. More comprehensive, end-to-end deep learning models to combine multiple sources of data to generate accurate final protein structures can be developed to automatically and accurately reconstruct protein structures from the data. 

Moreover, it is important to integrate cryo-EM based deep learning methods of reconstructing protein structures with the advanced methods developed in the field of protein structure prediction. The structural models directly reconstructed from cryo-EM data by deep learning  generally have correct overall topology, but the reconstructed models may not satisfy physicochemical restraints such as bond length and bond angles and not have all the molecular details (e.g., the precise location of all side chain atoms) \cite{cascaded_cnn, deepproligand}. Linking the atoms of amino acids identified from the density maps into full peptide chains consistent with protein sequences and physicalchemical restraints is still challenging.  However, the modeling techniques such as protein structure refinement and molecular dynamics to fix these problems have been established for protein structure prediction \cite{alphafold}. Some methods such as CR-I-TASSER have started to integrate the two kinds of technologies. More synergistic integration of the two are needed to generate high-quality realistic protein structures from cryo-EM data. 

The development of high-quality deep learning models to reconstruct protein structures from cryo-EM density maps critically depends on the availability of sufficient high-quality training data. Curating a large amount of high-quality training and test data is challenging and time consuming, but often receives little attention. Currently, there are few well-curated data sets available for training and evaluating deep learning models in the field. Therefore, more effort needs to be devoted to creating such data sets and make them to publicly available for the community to use. 

\section{Conclusion}

A number of useful deep learning models have been developed to reconstruct protein structures from cryo-EM density maps, demonstrating deep learning is a promising technology to further push the frontier of applying cryo-EM technology to determine protein structures. As the deep learning field is evolving very fast, many more state-of-the-art deep learning architectures (e.g., AlphaFold2-like models and transformers) have yet to be applied to further advance the emerging field. More sophisticated deep learning methods need to be developed to seemlessly integrate cryo-EM data with other complementary data such as predicted protein structures, protein sequences, and template structures to further improve cryo-EM-based structure determination. A synergistic integration of cryo-EM based protein structure determination techniques and latest protein structure prediction techniques is also important for generating highly accurate native-like protein structures.  To speed up the development, more effort is need to create a large amount of high-quality cryo-EM training and test data for the community to use.   

\section{Conflict of interest statement}
The authors declare that there is no conflict of interest. 

\section{Acknowledgements}
This work was supported in part by Department of Energy grants (DE-AR0001213, DE-SC0020400, and DE-SC0021303), two NSF grants (DBI1759934 and IIS1763246), and NIH grants (R01GM093123 and R01GM146340).


\begin{thebibliography}{GMS98} \item[] \hskip-\leftmargin \begin{minipage}{\textwidth} 
Papers of particular interest, published within the period of review, have been highlighted as:\\

$\ast$ of special interest\\
$\ast \ast$ of outstanding interest
\end{minipage} 
\bigskip 

\bibitem {alphafold}  $\ast \ast$ \color{Black}Jumper J, Evans R, Pritzel A, Green T, Figurnov M, Ronneberger O, Tunyasuvunakool K, Bates R, Zídek A, Potapenko A, et al.: Highly accurate protein structure prediction with AlphaFold. Nature 2021, https://doi.org/10.1038/ s41586-021-03819-2. \\
\color{black} \textit{Highly accurate deep neural network based system that predicts protein's 3D structure form its amino acid sequence.}


\bibitem{EMDB} \color{MidnightBlue} Catherine L. Lawson, Ardan Patwardhan, Matthew L. Baker, Corey Hryc, Eduardo Sanz Garcia, Brian P. Hudson, Ingvar Lagerstedt, Steven J. Ludtke, Grigore Pintilie, Raul Sala, John D. Westbrook, Helen M. Berman, Gerard J. Kleywegt, Wah Chiu, EMDataBank unified data resource for 3DEM, Nucleic Acids Research, Volume 44, Issue D1, 4 January 2016, Pages D396–D403, https://doi.org/10.1093/nar/gkv1126

\bibitem {resolution_revolution} \color{MidnightBlue} Werner Kühlbrandt : The Resolution Revolution. Science 2014, 10.1126/science.1251652 , https://www.science.org/doi/abs/10.1126/science.1251652 s41586-021-03819-2

\bibitem[]{deepproligand} \color{MidnightBlue} Giri, N., Cheng, J. (2022). A Deep Learning Bioinformatics Approach to Modeling Protein-Ligand Interaction with cryo-EM Data in 2021 Ligand Model Challenge. bioRxiv.

\bibitem[]{rosettafold}  \color{MidnightBlue} Baek, M., DiMaio, F., Anishchenko, I., Dauparas, J., Ovchinnikov, S., Lee, G. R., ...  Baker, D. (2021). Accurate prediction of protein structures and interactions using a three-track neural network. Science, 373(6557), 871-876. \\

\bibitem {emnuss} \color{MidnightBlue} He, J.,  Huang, S. Y. (2021). EMNUSS: a deep learning framework for secondary structure annotation in cryo-EM maps. Briefings in bioinformatics, 22(6), bbab156.

\bibitem[]{structure_generator} \color{MidnightBlue} Li, Po-Nan, Saulo HP de Oliveira, Soichi Wakatsuki, and Henry van den Bedem. "Sequence-guided protein structure determination using graph convolutional and recurrent networks." In 2020 IEEE 20th international conference on bioinformatics and bioengineering (BIBE), pp. 122-127. IEEE, 2020.

\bibitem[]{emap2sec} \color{MidnightBlue} Maddhuri Venkata Subramaniya, Sai Raghavendra, Genki Terashi, and Daisuke Kihara. "Protein secondary structure detection in intermediate-resolution cryo-EM maps using deep learning." Nature methods 16, no. 9 (2019): 911-917.

\bibitem[]{aanchor} \color{MidnightBlue} Rozanov, Mark, and Haim J. Wolfson. "AAnchor: CNN guided detection of anchor amino acids in high resolution cryo-EM density maps." In 2018 IEEE international conference on bioinformatics and biomedicine (BIBM), pp. 88-91. IEEE, 2018.
 
\bibitem[]{cascaded_cnn} \color{black} $\ast$ $\ast$ Si, Dong, Spencer A. Moritz, Jonas Pfab, Jie Hou, Renzhi Cao, Liguo Wang, Tianqi Wu, and Jianlin Cheng. "Deep learning to predict protein backbone structure from high-resolution cryo-EM density maps." Scientific reports 10, no. 1 (2020): 1-22 \\
\textit{Among the first deep learning methods that accurately predict C$\alpha$ positions along the protein's backbone from cryo-EM density maps automatically.}

\bibitem[]{cnn_based} \color{MidnightBlue} Li, Rongjian, Dong Si, Tao Zeng, Shuiwang Ji, and Jing He. "Deep convolutional neural networks for detecting secondary structures in protein density maps from cryo-electron microscopy." In 2016 IEEE International Conference on Bioinformatics and Biomedicine (BIBM), pp. 41-46. IEEE, 2016.

\bibitem[]{haruspex} \color{black} $\ast$ Mostosi, Philipp, Hermann Schindelin, Philip Kollmannsberger, and Andrea Thorn. "Haruspex: a neural network for the automatic identification of oligonucleotides and protein secondary structure in cryo‐electron microscopy maps." Angewandte Chemie International Edition 59, no. 35 (2020): 14788-14795. \\
\textit{Identifies secondary structures and nucleotides with high precision and recall.}


\bibitem{deeptracer} \color{black} $\ast \ast$ Pfab, Jonas, Nhut Minh Phan, and Dong Si. "DeepTracer for fast de novo cryo-EM protein structure modeling and special studies on CoV-related complexes." Proceedings of the National Academy of Sciences 118, no. 2 (2021): e2017525118. \\
\textit{Builds 3D protein structure automatically and accurately from cryo-EM and amino acid sequence.}

\bibitem[]{deeptracer_id} \color{black} $\ast$ Chang, Luca, Fengbin Wang, Kiernan Connolly, Hanze Meng, Zhangli Su, Virginija Cvirkaite-Krupovic, Mart Krupovic, Edward H. Egelman, and Dong Si. "DeepTracer ID: De Novo Protein Identification from Cryo-EM Maps." bioRxiv (2022).\\
\textit{Predicts backbone structures using DeepTracer and searches them against AlphaFoldDB to refine the models. }


\bibitem{cr_i_tasser} \color{MidnightBlue} Zhang, Xi, Biao Zhang, Peter L. Freddolino, and Yang Zhang. "CR-I-TASSER: assemble protein structures from cryo-EM density maps using deep convolutional neural networks." Nature Methods 19, no. 2 (2022): 195-204.

\bibitem[]{embuild} \color{MidnightBlue} He, Jiahua, Peicong Lin, Ji Chen, Hong Cao, and Sheng-You Huang. "Model building of protein complexes from intermediate-resolution cryo-EM maps with deep learning-guided automatic assembly." Nature Communications 13, no. 1 (2022): 1-16.

\bibitem[]{emnuss} \color{MidnightBlue} He, Jiahua, and Sheng-You Huang. "EMNUSS: a deep learning framework for secondary structure annotation in cryo-EM maps." Briefings in bioinformatics 22, no. 6 (2021): bbab156.


\bibitem[]{cryo_drgn} \color{black} $\ast$  Zhong, Ellen D., Tristan Bepler, Bonnie Berger, and Joseph H. Davis. "CryoDRGN: reconstruction of heterogeneous cryo-EM structures using neural networks." Nature methods 18, no. 2 (2021): 176-185. \\
\textit{Classifies particle images in cryo-EM using variational autoencoder-decoder architecture.}


\bibitem{e2gmm} \color{MidnightBlue} Chen, Muyuan, and Steven J. Ludtke. "Deep learning-based mixed-dimensional Gaussian mixture model for characterizing variability in cryo-EM." Nature methods 18, no. 8 (2021): 930-936.

\bibitem{CDAE} \color{MidnightBlue} Lei, Houchao, and Yang Yang. "CDAE: a cascade of denoising autoencoders for noise reduction in the clustering of single-particle cryo-EM images." Frontiers in genetics 11 (2021): 627746.

\bibitem[]{map_denoising}  \color{MidnightBlue} Kimanius, Dari, Gustav Zickert, Takanori Nakane, Jonas Adler, Sebastian Lunz, C-B. Schönlieb, Ozan Öktem, and Sjors HW Scheres. "Exploiting prior knowledge about biological macromolecules in cryo-EM structure determination." IUCrJ 8, no. 1 (2021): 60-75.

\bibitem{em_fold} \color{MidnightBlue} Lindert, Steffen, Nathan Alexander, Nils Wötzel, Mert Karakaş, Phoebe L. Stewart, and Jens Meiler. "EM-fold: de novo atomic-detail protein structure determination from medium-resolution density maps." Structure 20, no. 3 (2012): 464-478.

\bibitem{gorgon} \color{MidnightBlue} Baker, Matthew L., Sasakthi S. Abeysinghe, Stephen Schuh, Ross A. Coleman, Austin Abrams, Michael P. Marsh, Corey F. Hryc, Troy Ruths, Wah Chiu, and Tao Ju. "Modeling protein structure at near atomic resolutions with Gorgon." Journal of structural biology 174, no. 2 (2011): 360-373.

\bibitem{rosetta} \color{MidnightBlue} DiMaio, Frank, Andrew Leaver-Fay, Phil Bradley, David Baker, and Ingemar André. "Modeling symmetric macromolecular structures in Rosetta3." PloS one 6, no. 6 (2011): e20450.

\bibitem{pathwalking} \color{MidnightBlue} Chen, Muyuan, Philip R. Baldwin, Steven J. Ludtke, and Matthew L. Baker. "De Novo modeling in cryo-EM density maps with Pathwalking." Journal of structural biology 196, no. 3 (2016): 289-298.

\bibitem{mainmastseg} \color{MidnightBlue} Terashi, Genki, Yuki Kagaya, and Daisuke Kihara. "MAINMASTseg: automated map segmentation method for cryo-EM density maps with symmetry." Journal of chemical information and modeling 60, no. 5 (2020): 2634-2643.

\bibitem{mainmast} \color{MidnightBlue} Terashi, Genki, and Daisuke Kihara. "De novo main-chain modeling for EM maps using MAINMAST." Nature communications 9, no. 1 (2018): 1-11.

\bibitem{phenix} \color{MidnightBlue} Liebschner, Dorothee, Pavel V. Afonine, Matthew L. Baker, Gábor Bunkóczi, Vincent B. Chen, Tristan I. Croll, Bradley Hintze et al. "Macromolecular structure determination using X-rays, neutrons and electrons: recent developments in Phenix." Acta Crystallographica Section D: Structural Biology 75, no. 10 (2019): 861-877.


\bibitem{rennsh} \color{MidnightBlue} Ma, Lingyu, Marco Reisert, and Hans Burkhardt. "RENNSH: A Novel$\alpha$-Helix Identification Approach for Intermediate Resolution Electron Density Maps." IEEE/ACM Transactions on Computational Biology and Bioinformatics 9, no. 1 (2011): 228-239.

\bibitem{sselearner} \color{MidnightBlue} Si, Dong, Shuiwang Ji, Kamal Al Nasr, and Jing He. "A machine learning approach for the identification of protein secondary structure elements from electron cryo‐microscopy density maps." Biopolymers 97, no. 9 (2012): 698-708.

\bibitem{cryogan} \color{MidnightBlue} Gupta, Harshit, Michael T. McCann, Laurene Donati, and Michael Unser. "CryoGAN: a new reconstruction paradigm for single-particle cryo-EM via deep adversarial learning." IEEE Transactions on Computational Imaging 7 (2021): 759-774.

\bibitem{cnn} \color{MidnightBlue} Rawat, Waseem, and Zenghui Wang. "Deep convolutional neural networks for image classification: A comprehensive review." Neural computation 29, no. 9 (2017): 2352-2449.


\bibitem {unet} \color{black}  $\ast$ Ronneberger, Olaf, Philipp Fischer, and Thomas Brox. "U-net: Convolutional networks for biomedical image segmentation." In International Conference on Medical image computing and computer-assisted intervention, pp. 234-241. Springer, Cham, 2015. \\
\textit{U-Net, a widely used architecture to classify/segment pixels in medical images.}

\bibitem{cryo_review} \color{black} $\ast$ Si, Dong, Andrew Nakamura, Runbang Tang, Haowen Guan, Jie Hou, Ammaar Firozi, Renzhi Cao, Kyle Hippe, and Minglei Zhao. "Artificial intelligence advances for de novo molecular structure modeling in cryo‐electron microscopy." Wiley Interdisciplinary Reviews: Computational Molecular Science 12, no. 2 (2022): e1542\\
\textit{A systematic review for AI methods in cryo-EM, covering implementation of AI in different stages of cryo-EM workflow.}

\bibitem{transformer} \color{black} $\ast$ $\ast$ Vaswani, Ashish, Noam Shazeer, Niki Parmar, Jakob Uszkoreit, Llion Jones, Aidan N. Gomez, Łukasz Kaiser, and Illia Polosukhin. "Attention is all you need." Advances in neural information processing systems 30 (2017). \\
\textit{Deep learning model that uses self-attention module to learn and identify relationships in data. Originally used in the fields of natural language processing and computer vision. }

\bibitem{se3} \color{black} $\ast$ $\ast$ Fuchs, Fabian, Daniel Worrall, Volker Fischer, and Max Welling. "Se (3)-transformers: 3d roto-translation equivariant attention networks." Advances in Neural Information Processing Systems 33 (2020): 1970-1981. \\
\textit{ SE(3)-Transformer, a variant of the self-attention module for 3D point clouds, which is equivariant under continuous 3D roto-translations.}

\bibitem{deepcryopicker} \color{MidnightBlue} Al-Azzawi, Adil, Anes Ouadou, Highsmith Max, Ye Duan, John J. Tanner, and Jianlin Cheng. "DeepCryoPicker: fully automated deep neural network for single protein particle picking in cryo-EM." BMC bioinformatics 21, no. 1 (2020): 1-38.

\bibitem[]{orf3a} \color{MidnightBlue} Kern, David M et al. “Cryo-EM structure of SARS-CoV-2 ORF3a in lipid nanodiscs.” Nature structural and molecular biology vol. 28,7 (2021): 573-582. doi:10.1038/s41594-021-00619-0

\bibitem{PDB} \color{MidnightBlue} Stephen K Burley, Charmi Bhikadiya, Chunxiao Bi et al. RCSB Protein Data Bank: powerful new tools for exploring 3D structures of biological macromolecules for basic and applied research and education in fundamental biology, biomedicine, biotechnology, bioengineering and energy sciences, Nucleic Acids Research, Volume 49, Issue D1, 8 January 2021, Pages D437–D451, https://doi.org/10.1093/nar/gkaa1038

\bibitem{gcn} \color{MidnightBlue} Kipf, Thomas N., and Max Welling. "Semi-supervised classification with graph convolutional networks." arXiv preprint arXiv:1609.02907 (2016).

\bibitem[]{lstm} \color{MidnightBlue} Sherstinsky, Alex. "Fundamentals of recurrent neural network (RNN) and long short-term memory (LSTM) network." Physica D: Nonlinear Phenomena 404 (2020): 132306.

\bibitem[]{unet++} \color{black} $\ast$ Zhou, Zongwei, Md Mahfuzur Rahman Siddiquee, Nima Tajbakhsh, and Jianming Liang. "Unet++: A nested u-net architecture for medical image segmentation." In Deep learning in medical image analysis and multimodal learning for clinical decision support, pp. 3-11. Springer, Cham, 2018. \\
\textit{Nested U-Net, also known as U-Net ++, architecture that performs better than vanilla U-Net in segmentation tasks.}

\bibitem[]{machine_learning} \color{MidnightBlue} Greener, Joe G., Shaun M. Kandathil, Lewis Moffat, and David T. Jones. "A guide to machine learning for biologists." Nature Reviews Molecular Cell Biology 23, no. 1 (2022): 40-55.

\bibitem[]{deep_learning_medical} \color{MidnightBlue} Esteva, Andre, Katherine Chou, Serena Yeung, Nikhil Naik, Ali Madani, Ali Mottaghi, Yun Liu, Eric Topol, Jeff Dean, and Richard Socher. "Deep learning-enabled medical computer vision." NPJ digital medicine 4, no. 1 (2021): 1-9.

\bibitem{i_tasser} \color{MidnightBlue} Yang, Jianyi, Renxiang Yan, Ambrish Roy, Dong Xu, Jonathan Poisson, and Yang Zhang. "The I-TASSER Suite: protein structure and function prediction." Nature methods 12, no. 1 (2015): 7-8.

\bibitem{attention_survey} \color{MidnightBlue} Guo, Meng-Hao, Tian-Xing Xu, Jiang-Jiang Liu, Zheng-Ning Liu, Peng-Tao Jiang, Tai-Jiang Mu, Song-Hai Zhang, Ralph R. Martin, Ming-Ming Cheng, and Shi-Min Hu. "Attention mechanisms in computer vision: A survey." Computational Visual Media (2022): 1-38.

\bibitem{glycoprotein} \color{MidnightBlue} Gui M, Song W, Zhou H, Xu J, Chen S, Xiang Y, Wang X. Cryo-electron microscopy structures of the SARS-CoV spike glycoprotein reveal a prerequisite conformational state for receptor binding. Cell Res. 2017 Jan;27(1):119-129. doi: 10.1038/cr.2016.152. Epub 2016 Dec 23. PMID: 28008928; PMCID: PMC5223232.

\bibitem{chimerax} \color{MidnightBlue} Pettersen, Eric F., Thomas D. Goddard, Conrad C. Huang, Elaine C. Meng, Gregory S. Couch, Tristan I. Croll, John H. Morris, and Thomas E. Ferrin. "UCSF ChimeraX: Structure visualization for researchers, educators, and developers." Protein Science 30, no. 1 (2021): 70-82

\bibitem[]{eman2} \color{MidnightBlue} Bell JM, Chen M, Durmaz T, Fluty AC, Ludtke SJ. New software tools in EMAN2 inspired by EMDatabank map challenge. J Struct Biol. 2018 Nov;204(2):283-290. doi: 10.1016/j.jsb.2018.09.002. Epub 2018 Sep 4. PMID: 30189321; PMCID: PMC6163079.

\bibitem[]{vesper} \color{MidnightBlue} Alnabati, Eman, Genki Terashi, and Daisuke Kihara. "Protein Structural Modeling for Electron Microscopy Maps Using VESPER and MAINMAST." Current Protocols 2, no. 7 (2022): e494.


\bibitem[]{situs} \color{MidnightBlue} Wriggers, Willy. "Using Situs for the integration of multi-resolution structures." Biophysical reviews 2, no. 1 (2010): 21-27.


\bibitem{geometric_deeplearning} \color{black} $\ast$ $\ast$ Bronstein, Michael M., Joan Bruna, Taco Cohen, and Petar Veličković. "Geometric deep learning: Grids, groups, graphs, geodesics, and gauges." arXiv preprint arXiv:2104.13478 (2021).\\
\textit{A (proto-) book on geometric deep learning about representational learning architectures and exploiting the symmetries of data therein.}

\end{thebibliography}


\end{document}